\documentclass[twocolumn]{aa} 
\usepackage{multirow}
\usepackage{amsmath}
\usepackage{amsfonts}
\usepackage{amssymb}
\usepackage{mathrsfs}
\usepackage{verbatim}
\usepackage{siunitx}

\usepackage{graphicx}
\usepackage{subfig}
\usepackage{color}      
\usepackage{natbib}
\begin{document}

\title{Nitrile versus isonitrile adsorption at interstellar grains surfaces: \\ I - Hydroxylated surfaces}

   \author{     M. Bertin, \inst{1} 
                   M. Doronin, \inst{1,2}   
                J.-H. Fillion,  \inst{1} 
                X. Michaut,  \inst{1} 
                L. Philippe,  \inst{1}
                        M. Lattelais, \inst{2}          
                        A. Markovits, \inst{2}
                        F. Pauzat, \inst{2}
                        Y. Ellinger, \inst{2}    
                J.-C. Guillemin  \inst{3} 
          }

   \offprints{M. Bertin}

          \institute{LERMA, Sorbonne Universit\'es, UPMC Univ. Paris 06, Observatoire de Paris, PSL Research University, CNRS, F-75252, Paris, France\\
                   \email{mathieu.bertin@upmc.fr}
         \and
                Sorbonne Universit\'{e}s, UPMC Univ. Paris 06, UMR - CNRS 7616, Laboratoire de Chimie Th\'{e}orique, F-75252 Paris, France\\
                \email{pauzat@lct.jussieu.fr} 
        \and 
                Institut des Sciences Chimiques de Rennes, Ecole Nationale Sup\'erieure de Chimie de Rennes, UMR - CNRS 6226,  F-35708, Rennes, France \\
                \email {jean-claude.guillemin@ensc-rennes.fr}   
        }

   \date{Received ???? , 2015; accepted ???? , 2015}

  
  \abstract
   {   Almost 20\% of the $\sim200$ different species detected in the interstellar and circumstellar media 
   present a carbon atom linked to nitrogen by a triple bond. Among these 37 molecules, 30 are nitrile R-CN compounds, the remaining seven belonging to the isonitrile R-NC family.  How these species behave in presence of the grain surfaces is still an open question. } 
   {In this contribution we investigate whether the difference between  nitrile and isonitrile functional groups may induce differences in the adsorption energies of the related  isomers at the surfaces of interstellar grains of different nature and morphologies. }
   {The question was addressed by means of a concerted experimental and theoretical study of the adsorption energies of CH$_3$CN and CH$_3$NC on the surface water ice and silica. 
  The experimental determination of the molecule - surface interaction energies was  carried out  
   using temperature programmed desorption (TPD) under an ultra-high vacuum (UHV)  between 70 and 160 K. 
   Theoretically, the question was  addressed 
   using first principle periodic density functional theory (DFT) to represent  the organized solid support. } 
   {The most stable isomer (CH$_3$CN) 
   interacts more efficiently with the solid support than the higher energy isomer (CH$_3$NC) for water ice and silica. 
 Comparing with the HCN and HNC pair of isomers, the simulations show an opposite behaviour, in which isonitrile HNC are more strongly adsorbed than nitrile HCN provided that hydrogen bonds are compatible with the nature of the model surface. } 
   {The present study confirms that  the strength of the molecule surface interaction between isomers is not related to their intrinsic stability but instead to  their respective ability to generate different types of hydrogen bonds. 
   Coupling TPD to first principle simulations is
      a powerful method for investigating the possible role of interstellar surfaces  in the release of organic species from grains, depending on the environment.}

   \keywords{Astrochemistry; ISM: molecules -- ISM: abundances; Methods: laboratory -- Methods: numerical               }
   \titlerunning{Nitrile versus Isonitrile adsorption }  
     \authorrunning{Bertin et al.}  
   \maketitle
%
\section{Introduction}

Among the $\sim$200 molecules detected in the interstellar medium (ISM) gas phase, ({\it http://www.astro.uni-koeln.de/cdms/molecules}), about one-fifth consists in nitriles R-CN and isonitriles R-NC molecules.  Considering purely organic  species, two different families can be identified  that cover the two classes of isomers: conjugated molecules in which R contains, or does not contain, a system of bonds that can be delocalized onto the CN triple bond. For the unsaturated species, the biggest family is unanmbiguously the cyanopolyyne family, which is formally derived by adding carbon atoms or C$_2$ fragments to HCN;  the second family  is formally obtained by replacing the H atom in HCN and HNC by  saturated hydrocarbon fragments. The simplest common ancestor of both series is the 
CN radical, which is the second species ever detected in the ISM (McKellar 1940; Adams 1941; Jeffers et al. 1970). About the same time, HCN (Snyder \& Buhl 1971) and CH$_3$CN (Solomon et al 1971) were detected in the ISM together with HC$_3$N (Turner 1971 ). The identification of HNC followed soon after (Snyder \& Buhl 1971;  Zuckerman et al 1972). The corresponding isonitrile CH$_3$NC was detected 13 years later (Cernicharo et al 1988) towards Sgr B2 and it took 25 more years to confirm the presence of this species in the Horsehead PDR (Gratier et al 2013). 
 
The abundance ratios between nitrile and isonitrile isomers have largely been used to constrain the chemical models proposed to account for their presence in the various objects in which they were observed. The HCN and HNC couple has been the most studied since, according to radio astronomy,  gas-phase observations,  the ratio varies over a large range, namely, from values close to unity in cold molecular clouds (Irvine \& Schloerb 1984; Schilke et al 1992; Tennekes et al 2006) to several thousands in highly illuminated PDRs (Fuente et al 2003).  By contrast, the CH$_3$CN to CH$_3$NC abundance  ratio of  $\sim$50 appears remarkably stable (Irvine \& Schloerb 1984; Cernicharo et al 1988; Remijan et al 2005).

In this context, it should not be forgotten that radio observations are blind to molecules depleted on grains, 
and that gas-phase chemistry alone cannot account for the abundances of complex organic  molecules observed in the ISM.  Solid-gas reactions, and even chemical processes involving adsorbed partners,  have to be considered, making adsorption energies a key factor for the determination of real abundances and abundance ratios. 
In spite of a strong demand, very little quantitative data are available in regards to adsorption energies,  particularly concerning the nitrile and isonitrile families. 
This is mainly because of the difficulty of obtaining isonitriles (HNC or CH$_3$NC) in the laboratory and to the well-known lethal action of HCN. Consequently, we present a joint research mixing experiments and numerical simulations, focused on the determination of the adsorption energies of the first members of the nitrile and isonitrile families on laboratory analogues of widespread grain surfaces. It is clear, for the above reasons, that the experiments were limited to the isomers CH$_3$CN and CH$_3$NC.

Among the various grain models originally proposed (Greenberg 1976), such as ices, refractory minerals, and 
 carbonaceous dust particles, we selected in this first report the two hydroxylated components,  
namely, water ice and silica taken as a well-defined refractory model. 
 Water ices constitute a major part of the condensed matter in cold and dense regions of the ISM (Oberg et al. 2011; Boogert et al. 2011;2015). 
 Most of these ices are commonly believed to be in amorphous phase, although crystalline ices have also been detected in warmer regions such as the Kuiper belt (Jenniskens et al. 1995; Jewitt \& Luu 2004; Molinari et al. 1999). 
 Silicates in different forms and silica itself, whose presence in the ISM has been disputed for a long time, are now clearly identified in protostellar envelopes (Poteet et al. 2011) and protoplanetary disks (Juhasz et al. 2010; Sargent et al. 2009) using mid- and far-infrared spectra obtained by the Spitzer Space Telescope. 
 
 These two types of solids are recognized to have a potential impact on the distribution of the volatiles between the gas phase and dust grains.

The article is organized as follows. In Section 2 we present the experimental setup, method and results for adsorption energy determination of CH$_3$CN and CH$_3$NC on $\alpha$-quartz (0001), compact amorphous and crystalline water ice. In Section 3, we present the theoretical approach beginning by a brief outline of the methods used followed by the results of the calculations. In addition we include those results on HCN and HNC.
 In the last section, the theoretical and experimental results are compared and discussed\ with an emphasis on the possible implications on the observed abundances. We are not concerned here about the possible role of the grain surfaces in the formation of  nitriles and isonitriles.

\section{Experimental approach to adsorption energies}


\subsection{Experimental setup}

The experimental studies were performed in the setup SPICES (UPMC; Paris) under ultra-high vacuum conditions (base pressure of $\sim 1 \times 10^{-10}$ Torr). The solid substrates used in the studies, namely polycrystalline gold and monocrystalline $\alpha$-quartz (0001), are mounted on the cold tip of a rotatable closed-cycle helium cryostat. Resistive heating allows for the control of the substrates temperature between 10 and 300 K with a precision better than 0.1 K. Water substrates can be grown on either quartz or gold surfaces by exposing them \textit{in situ} at low temperature to H$_2$O vapor (Fluka, HPLC purity standard) through a dosing tube positioned a few millimeters in front of the sample. Morphology of the water ice substrate can be controlled by tuning the deposition temperature of the supporting surface: amorphous compact water ice is grown on surfaces kept at 100 K, while polycrystalline water ice is prepared at 150 K. The resulting water ice morphology is systematically checked by monitoring its thermal desorption during the experiments. 

Commercially available acetonitrile (CH$_3$CN, Sigma-Aldricht, 99\% purity) and and freshly synthesized methylisocyanide (CH$_3$NC)  are deposited on the chosen cold substrate (gold, quartz or water ice surface) following the same protocol as for the water ice growth. Each chemical product is further purified by several freeze-pump-thaw cycles prior to its introduction into the setup, and the purity is checked by mass spectrometry and infrared spectroscopy as detailed in the next sections. The CH$_3$CN or CH$_3$NC overlayer thickness is expressed in monolayers (ML), one monolayer corresponding to a saturated wetting molecular layer on the substrate. Our deposition method, which takes the pumping inertia of the molecules in the setup into account, gives a reproducibility of a few percent on the amount of deposited molecules.  

The physisorption of the CH$_3$CN and CH$_3$NC is experimentally studied by means of the temperature programmed desorption (TPD) technique. The sample is warmed up with a constant heating rate, while the desorbing species are detected as a function of the temperature by a quadripolar mass spectrometer (QMS Prisma; Pfeiffer). The signal of the desorbed molecules can be converted to a desorption flux, from which thermodynamical data such as adsorption energy can be extracted. An explanation of this procedure, together with the absolute coverage calibration from the TPD curves, can be found in section 2.3. Finally, the condensed molecular film can be probed after the ice preparation and during the warming up, using Fourier transform-reflection absorption infrared spectrocopy (FT-RAIRS). In practice, this is achieved by means of a Brucker vec22 spectrometer, with a maximum resolution of 1 cm$^{-1}$, coupled to the main vacuum chamber by differentially pumped KBr windows.

\begin{figure}[h!]
\begin{minipage}{\columnwidth} 
\centering

\includegraphics[scale=0.3, angle=0]{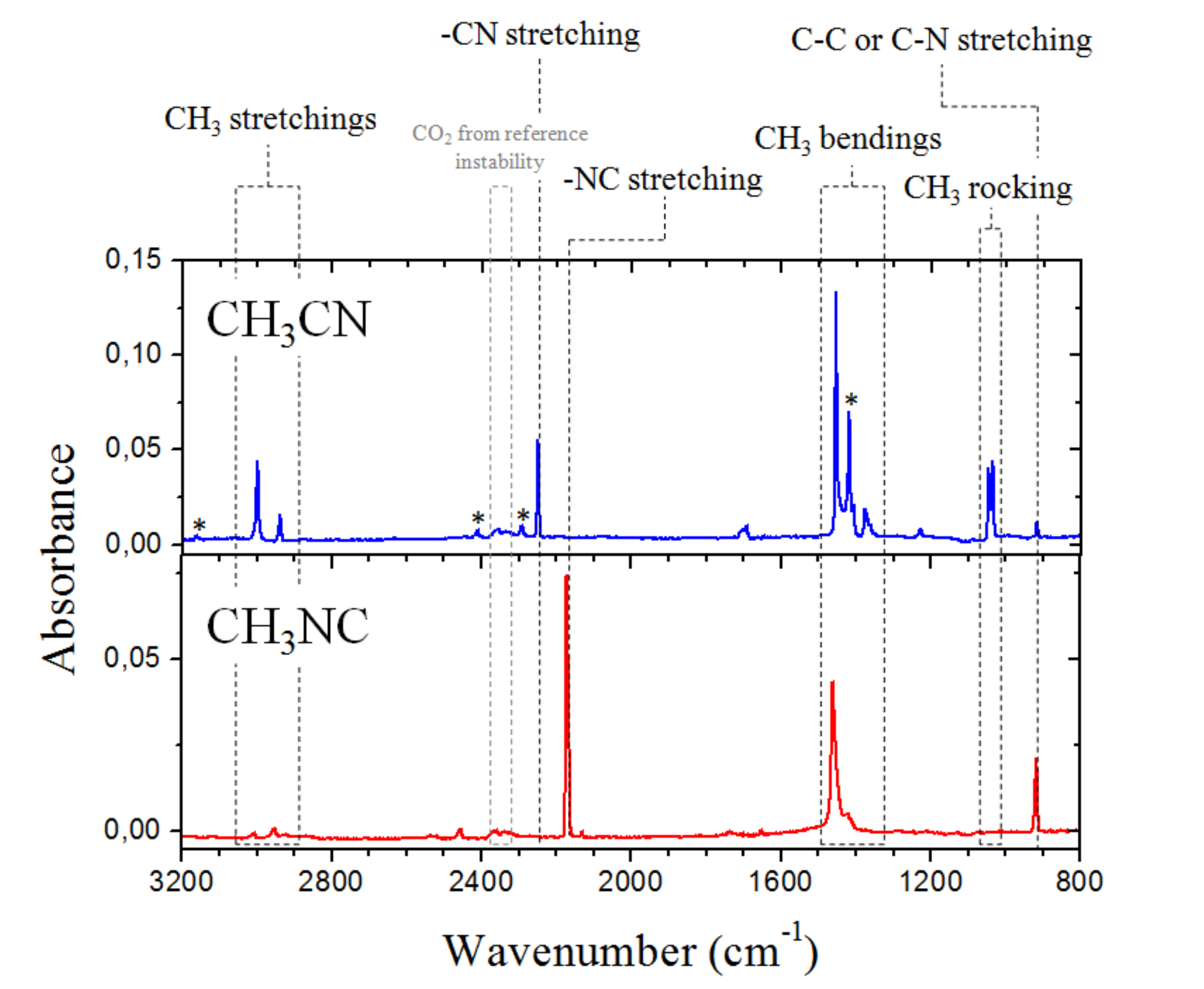} 
\end{minipage}
 
\caption{Reflection-absorption infrared spectra of thick CH$_3$CN ice (850 ML; upper panel) and of thick CH$_3$NC  ice (890 ML; lower panel), condensed at 90 K on polycrystalline gold. Indicated assignments are detailed in Table 1. Peaks denoted with (*) correspond to identified combination modes. Structures observed at $\approx$ 2350 cm$^{-1}$ are associated with a modification of the reference spectra due to a change of gaseous CO$_2$ concentration in the spectrometer during the experiment time.  } 
\end{figure}

\subsection{Pure thick ices of CH$_3$CN and CH$_3$NC}

\begin{table*}
\begin{center}
\caption{Vibration wavenumbers (in cm$^{-1}$) and mode assignments of gas phase and physisorbed multilayers of CH$_3$CN and CH$_3$NC in the 800 - 3200 cm$^{-1}$ range. \label{tab1} }
\begin{tabular}{ccccccc}     
\hline
\hline       
\multicolumn{3}{c}{Acetonitrile CH$_3$CN} &  \multicolumn{3}{c}{Methylisocyanide CH$_3$NC} &  \\
Gas Phase$^a$ &  Multilayers$^b$ & This work & Gas Phase$^c$ &  Multilayers$^d$ & This work & Assignment \\
\hline
3178 & 3161 & 3163 & - & - & - & $\nu$(--C$\equiv$N) + $\nu$(C--C)  \\
3009 & 3002 & 3000 & 3014 & no & 3008 & $\nu_{as}$(CH$_3$) \\
2954 & 2941 & 2939 & 2966 & no & 2953 & $\nu_{s}$(CH$_3$) \\
2417 & 2415 & 2411 & - & - & - & $\delta_{s}$(CH$_3$) + $\rho$(CH$_3$) \\2305  & 2289 & 2293 & - & - & - & $\delta_{s}$(CH$_3$) + $\nu$(C--C) \\
2266 & 2250 & 2251 & - & - & - & $\nu$(--C$\equiv$N) \\
- & - & - & 2166 & 2170 & 2172 & $\nu$(--N$\equiv$C) \\
1448 & 1455 & 1455 & 1467 & 1470 & 1459 & $\delta_{as}$(CH$_3$) \\
1410 & 1422 & 1419 & - & - & - & $\rho$(CH$_3$) + $\delta$(C--C$\equiv$N) \\
1390 & 1378 & 1378 & 1429 & no & 1426 & $\delta_{s}$(CH$_3$) \\
1041 & 1038 & 1036 & 1129 & no & no &  $\rho$(CH$_3$) \\
- & - & - & 945 & no & 918 & $\nu$(C--N) \\
920 & no & 917 & - & - & - & $\nu$(C--C) \\ 
\hline
\end{tabular}
\end{center}
$^a$, $^b$, $^c$ and $^d$ are from Parker et al (1957) Schaff \& Roberts (1999), NIST chemical webbook ({\it http://webbook.nist.gov/chemistry/}) and Murphy et al (2000), respectively; $\nu$ stands for stretching mode, $\delta$ for bending mode, $\rho$ for rocking mode, $_{as}$ for asymmetric mode, and $_s$ for symmetric mode. The word `no' indicates non-observed vibrations.
\end{table*}

\bigskip
Before experimentally studying adsorption and desorption of CH$_3$CN and CH$_3$NC from different surfaces, it is  important to study these two species as pure ices first. For both isomers, we chose the mass signal associated with the ionization of the intact molecule (C$_2$H$_3$N$^+$, m = 41 amu) for both dosing and TPD experiments. The two species present comparable mass spectra following electron-induced ionization at 70 eV ({\it http://webbook.nist.gov/chemistry/}), making mass spectrometry unable to distinguish the two isomers unambiguously. This could potentially represent an important issue, in particular for CH$_3$NC. This molecule,  in its liquid form is much less stable than its isomer CH$_3$CN. 
CH$_3$NC was prepared using the synthesis of Schuster et al. (1966) but with trioctylamine instead quinoline as the base, 
and had to be kept under nitrogen at low temperature ($< $ 273K) and protected from light. It is therefore important to verify whether the product we introduced in the vacuum chamber is solely CH$_3$NC or if the molecule was partially degraded or isomerized to 
CH$_3$CN or CH$_2$CNH during the deposition or the warming up. To this purpose, we systematically
carried out FT-RAIRS on the CH$_3$CN and CH$_3$NC ices after their growth and during the warming up.

In Fig. 1 we present infrared spectra of thick ices ($\sim$ 800 ML) of pure CH$_3$CN (upper panel) and pure  CH$_3$NC (lower panel), which are deposited on polycrystalline gold at 90 K. We obtain infrared spectra with a resolution of 2 cm$^{-1}$. In the case of the condensed multilayer of acetonitrile, we obtain IR spectrum very similar to what has been found by Schaff \& Roberts (1999). The more intense peaks are associated with normal vibration modes of the condensed molecules, while many less intense peaks are related to combination modes; see Figure 1. Less literature is available on infrared spectroscopy of condensed methylisocyanide at low temperature. We used a vibrational study of condensed CH$_3$NC in the 1000 - 2500 cm$^{-1}$ range by Murphy et al.(2000) and gas-phase spectroscopy values for the assignment of CH$_3$NC IR peaks. The assignment we propose for the observed vibrational peaks are summarized in Table 1.

In the case of condensed methylisocyanide CH$_3$NC  ice, all the peaks are attributed to a given CH$_3$NC vibration, thereby confirming the purity of the deposited ice. The clear difference in the stretching vibrations of the cyanide (2250 cm$^{-1}$) and isocyanide (2170 cm$^{-1}$) groups means that the two isomers can be finely differentiated by infrared spectroscopy. The spectra in figure 1 show no contribution of the other isomer in each pure ice. This is still true during warming up and desorption of the ices, demonstrating that the mass signal monitored from pure isomer ices is not polluted in situ by any isomerization process during the TPD experiment.

%
\begin{figure*}
\centering
\includegraphics
[width=12cm]{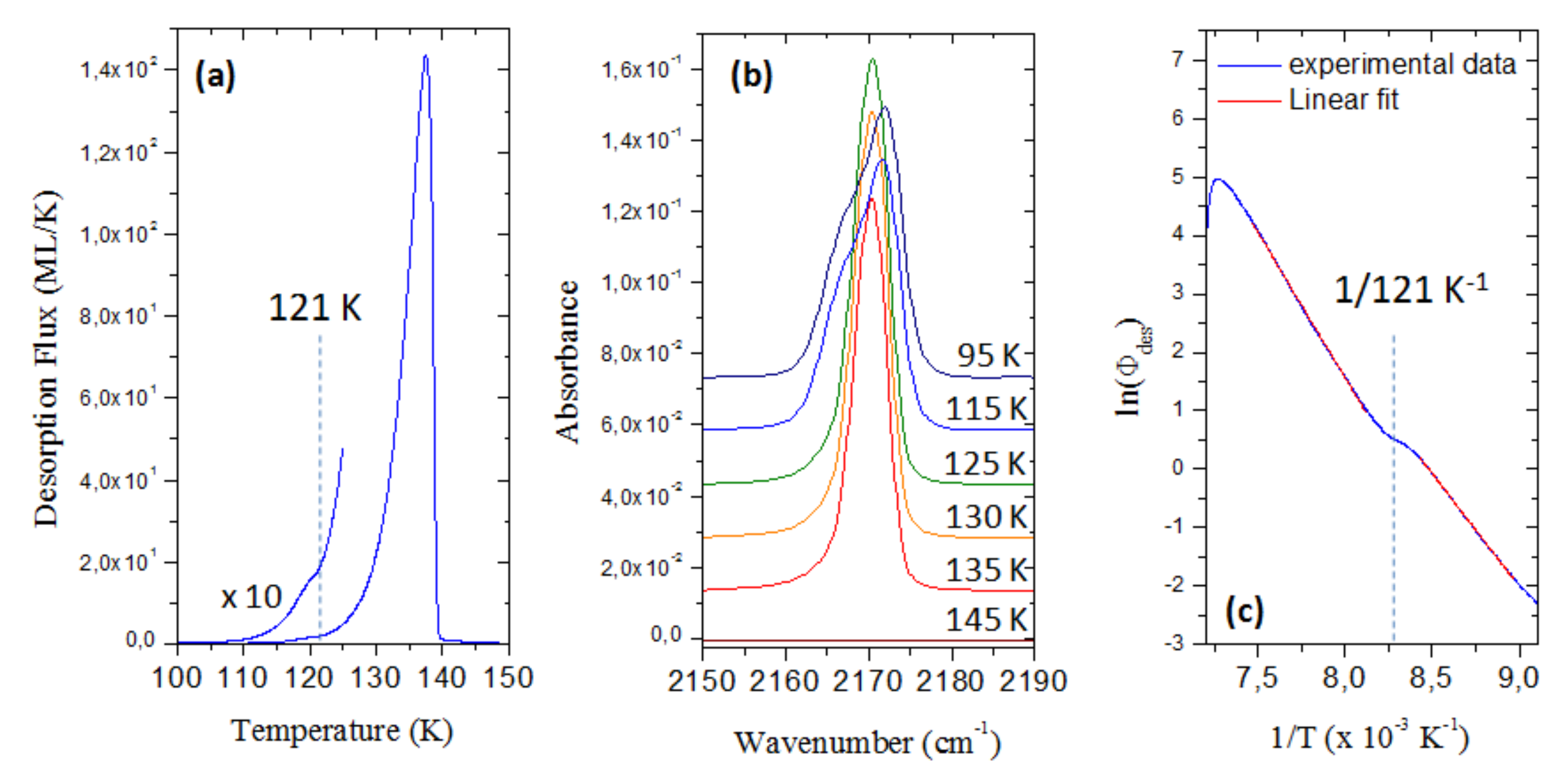} 
\caption{: (a) TPD curve, with a heating rate of 2 K/min, of $\sim$ 800 ML of pure CH$_3$NC deposited on polycrystalline gold at 90 K. The beginning of the desorption has been magnified by a factor of 10  (b) Infrared spectra in the region of the NC stretching vibration, during the warming up of the ice. (c) Plot of the logarithm of the desorption flux as a function of the inverse of the temperature. Lines are linear fits of the plot.}
\end{figure*}


Thermal desorption of the pure ices of each isomer was carried out to extract the relevant adsorption energy and prefactor of the Polanyi-Wigner law. Thermal desorption curves of multilayers of pure CH$_3$NC and pure 
CH$_3$CN on polycrystalline gold, using a heating rate of 2 K/min, are shown in Figures 2a and 3a, respectively. The multilayer desorption of CH$_3$NC presents a small shoulder at 121 K, which we attribute to a temperature-induced phase transition of the solid. This kind of phase transition from a disordered to a more structured solid is observed in several pure ices, of  which the most well-known case is the water ice crystallization at $\sim$150 K. The shoulder in the desorption curves is explained by different adsorption energies for each phase with a higher value for the most ordered phase in which molecules are usually more coordinated. The desorption of the disordered phase competes with its crystallization during the warming up, and the total  desorption flux observed is then the superimposition of the desorption of the disordered and  crystalline phases (Speedy et al 1996). This hypothesis is supported by the infrared follow-up of the NC stretching mode during the warming up, as shown in Figure 2b. Between 115 and 125 K, the vibrational peak gets narrower and more intense, whereas the amount of molecules on the surface is lower. This suggests a change of the local environment of the adsorbate, which modifies the dipole moment associated with the NC stretching vibration together with its bond strength. Such changes can be induced by a reorganization of the molecules in the solid, as has already been highlighted in the case of the C=O stretching modes in solid CH$_3$COOH and HCOOCH$_3$ (Bertin et al 2011; Modica \& Palumbo 2010)  or OH stretching modes in solid water (May et al  2011). However, this effect on the IR features can have different origins. Only its simultaneous observation with the shoulder in the TPD curve gives us enough clues to reach a conclusion about the existence of a phase transition. Interestingly, no phase transition was observed in solid CH$_3$CN either in the TPD curves or in the infrared spectra.

The adsorption energy of the multilayers of CH$_3$CN and CH$_3$NC can be derived from the TPD curves by considering that the desorption of the thick ices follows a zeroth order kinetic Polanyi-Wigner law. Indeed, zeroth order kinetics usually describes correctly the desorption of multilayers, since the amount of molecule at the vacuum-ice interface is kept constant during the desorption. In this case, the desorption flux can be expressed as
\begin{equation}
\Phi_{des}=-\frac{d\theta}{dT}=-\frac{d\theta}{dt}\frac{dt}{dT}=\frac{\nu}{\beta}\exp({-\frac{E_{ads}}{kT}})
,\end{equation}
\noindent where $\Phi_{des}$ is the thermal desorption flux, as measured by the mass spectrometer, $\theta$ is the molecule coverage ($\theta$ = 1 for a full compact monolayer on the substrate), $\beta$ is the heating rate, T is the temperature of the surface, k is the Boltzmann constant, $\nu$ is a prefactor, and E$_{ads}$ is the multilayer adsorption energy. This equation can be rewritten
\begin{equation}
\ln(\Phi_{des})=-\frac{E_{ads}}{k}\frac{1}{T}+\ln(\frac{\nu}{\beta})
.\end{equation}
Plotting the logarithm of the flux against 1/T results in a linear behaviour, in which the adsorption energy and prefactor can be derived from the slope and intercept, respectively. This linear trend is observed for CH$_3$NC and CH$_3$CN  (Figures 2c and 3b). In the case of CH$_3$NC, the plot consists of two different linear regimes, where each is associated with a different adsorption energy for the solid CH$_3$NC, which further confirms the hypothesis of a phase transition at 121 K. From linear fitting of the plots, 
we find for CH$_3$CN multilayer   E$_{ads}$ = 390 $\pm$ 10 meV and $\nu$ = 10$^{13 \pm0.5}$ s$^{-1}$. 
  For CH$_3$NC, we find E$_{ads}$  = 330 $\pm$ 10 meV and $\nu$ = 5$\times$10$^{12 \pm0.5}$ s$^{-1}$ 
  for the disordered phase and    E$_{ads}$  = 420 $\pm$ 10 meV and $\nu$ = 5$\times$10$^{16 \pm0.5}$ s$^{-1}$ for the crystalline phase.

%

\begin{figure}[h!]
\begin{minipage}{\columnwidth} 
\centering

\includegraphics[width=8cm]{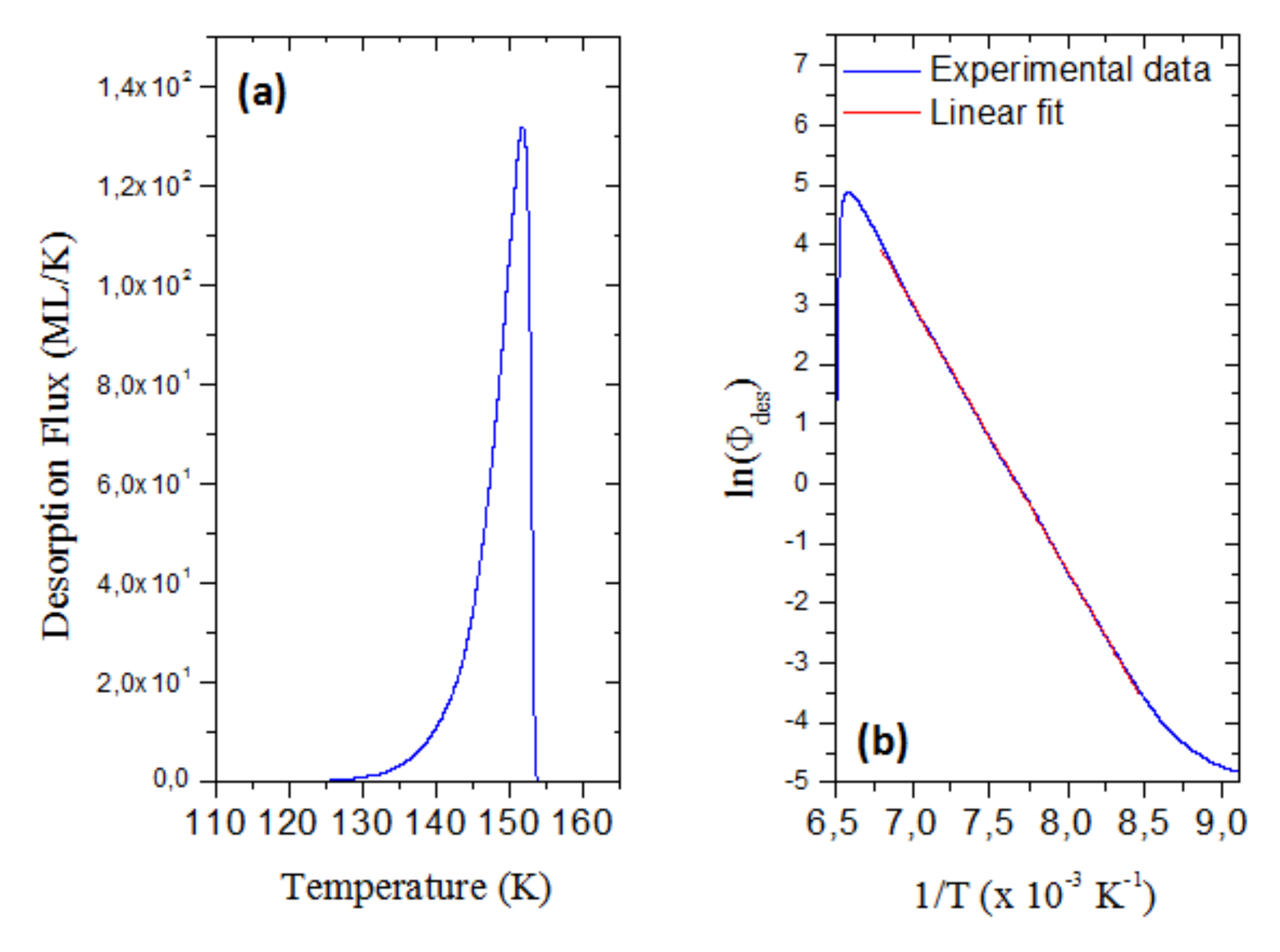} 
\end{minipage}
\caption{: (a) TPD curve, with a heating rate of 2 K/min, of $\sim$ 800 ML of pure CH$_3$CN deposited on polycrystalline gold at 90 K. (b) Plot of the logarithm of the desorption flux as a function of the inverse of the temperature. } 
\end{figure}

\subsection{Submonolayer of CH$_3$CN and CH$_3$NC deposited on model surfaces}

We studied the desorption of monolayer and submonolayer coverages of CH$_3$NC and CH$_3$CN on $\alpha$-quartz (0001) and compact amorphous and crystalline water ice to extract the adsorption energy of these isomers on the different substrates.

%

\begin{figure*}
\centering
\includegraphics
[width=15cm]{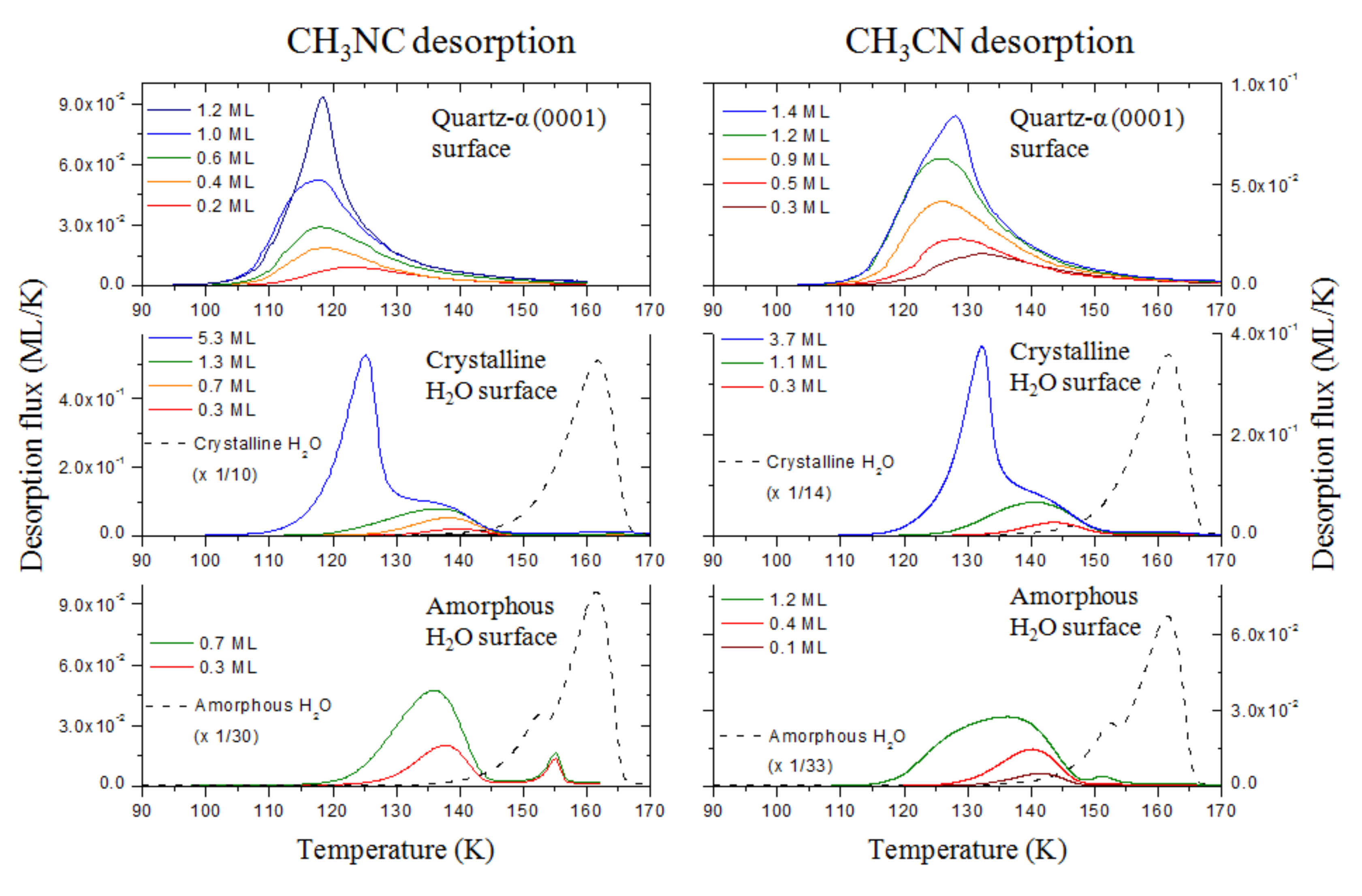} 
\caption{: TPD curves of several initial coverages of CH$_3$NC (left) and CH$_3$CN (right) adsorbed on $\alpha$-quartz (0001) surface,  and on crystalline and compact amorphous water ice surface. Applied heating rates are 12 K/min on quartz, and 10 K/min on water surfaces. TPD curves of multilayers ($\approx$ 50 ML) of crystalline and amorphous water ice are shown in dashed lines for comparison.  }
\end{figure*}

The associated TPD curves, carried out with a heating rate of 12 K/min and 10 K/min in the case of adsorption on water ices, are shown in Figure 4. The coverages indicated in the figure were calibrated for each surface. On the water ice surfaces, two desorption peaks can be observed. One at lower temperature is observed only for high exposure and, therefore, is related to the multilayer desorption. The second peak at higher temperature is associated with the desorption of the first monolayer, for which the adsorption energy is very different from that of the multilayer. The area of this second peak, when saturated, gives the absolute calibration for 1 ML. Finally, a weak desorption feature is seen at $\sim$ 155 K only in the case of adsorbates on compact amorphous water. This temperature corresponds to the supporting water ice crystallization, evidenced by a shoulder in the amorphous water TPD curve (dashed line). This is attributed to the phenomenon known as volcano effect: some adsorbates remain trapped in corrugations of the amorphous water ice and are suddenly released when crystallization of the water ice takes place (Ayotte et al 2001; Smith et al 1997). However, this contribution remains weak as the signal is dominated by the surface monolayer desorption. In the case of desorption from the quartz surface, only one desorption feature is observed, meaning that adsorption of the first layer and of the multilayer are not different enough to result in a clear separation in the desorption temperature. The transition between multilayer and monolayer adsorption can anyway be determined by observing a change in the desorption kinetics order  --- in this case from a zeroth order to a first order --- with lower exposure. This transition allows for the identification of the full monolayer coverage with a relative precision of the order of 10$\%$. The integral of the related TPD curve, proportional to the amount of the deposited molecule, can then be used to calibrate any initial coverage in terms of monolayers. Moreover, we simply divide the signal by this 1 ML integral to calibrate the TPD mass signal into desorption flux.  More details on this calibration procedure can be found in Doronin et al. 2015.

In order to derive the adsorption energy of each isomer from these surfaces, we use a method described in more detail in a previous work (Doronin et al. 2015). The desorption of the submonolayer coverage of molecules from the substrate is modelled using a first order approximation of the Polanyi-Wigner law. To obtain a more accurate description of the desorption, we consider a distribution of adsorption energies instead of a single value; this implicitly takes
different adsorption geometries and sites into account, including defects on the surface. The experimental desorption curves are then fitted using the following equation:
\begin{equation}
\Phi_{des}(T)=\frac{\nu}{\beta}\sum_i\theta_i(T)\exp({-\frac{E_{i}}{kT}})
,\end{equation}
\noindent where E{\it $_i$}  is the adsorption energy associated with site {\it i} and $\theta_i$ is the coverage of the molecules for  site {\it i}. In this case, we make the approximation of a prefactor $\nu,$ which does not depend on the adsorption site. The TPD curves obtained experimentally are fitted with this law using a simple sampling of the adsorption energy in the 350 - 550 meV range, where the only free parameter is the initial coverage of each site  {\it i} and taking an arbitrary initially value for $\nu$ (in this case, the initial arbitrary value chosen was $10^{15}$ $s^{-1}$). The result gives the initial amount of molecules adsorbed with the adsorption energy E{\it $_i$}, which is the adsorption energy distribution on the surface for a given initial coverage. The value of the prefactor $\nu$ needs however to be independently determined since any value results in appropriate fitting of the experimental data, but with a different adsorption energy distribution. For this purpose, the fitting procedure is applied to three TPD curves of the same initial coverage, but performed with three different heating rates.  An example of the procedure is illustrated in Figure 5 for CH$_3$CN adsorption on quartz. For a given value of $\nu$, the extracted adsorption energy distributions from the three curves is shifted in energy. 
%
\begin{figure*}
\centering
\includegraphics
[width=12cm]{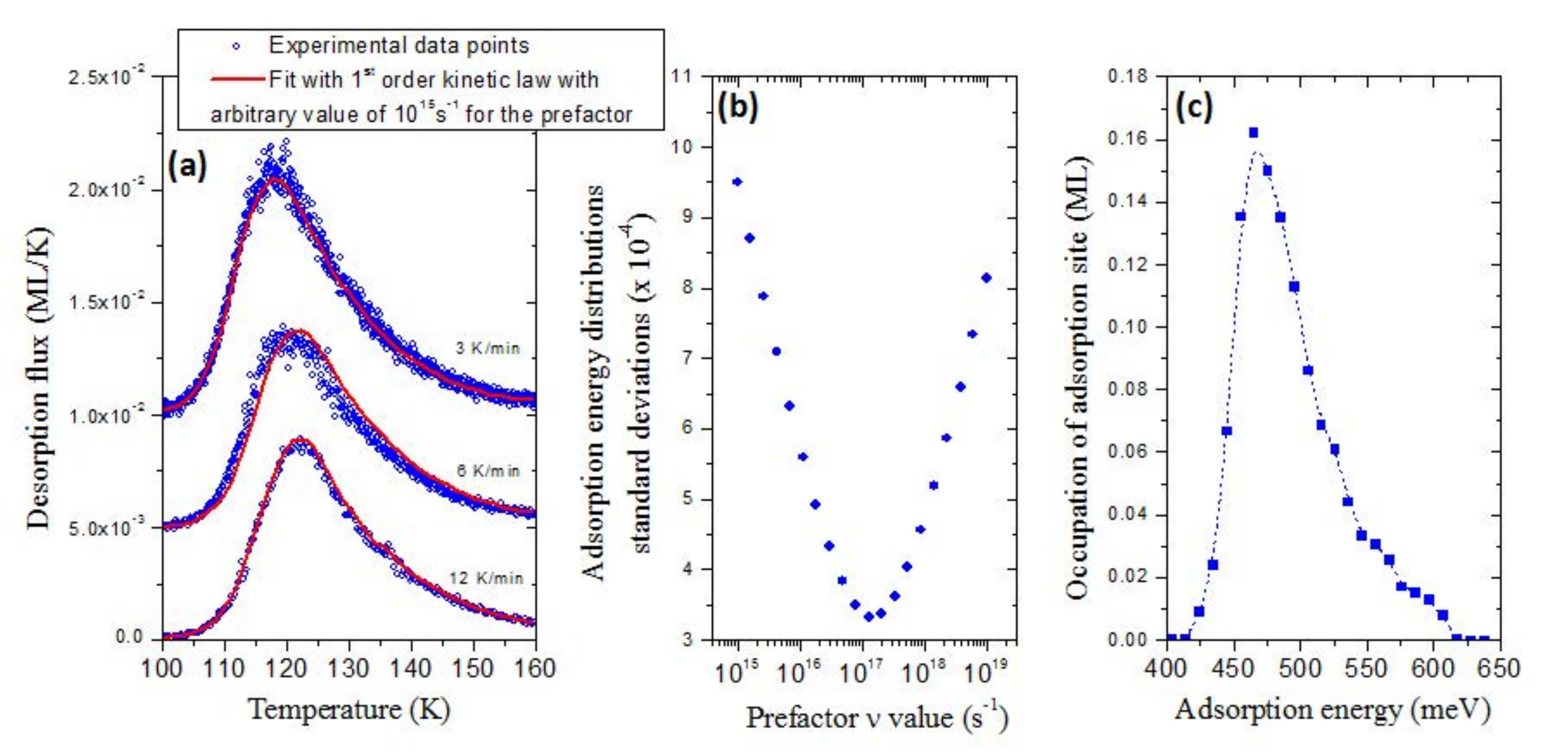} 
\caption{: (a) Set of TPD curves of 0.2 ML of CH$_3$CN adsorbed on $\alpha$-quartz (0001) performed at 3 K/min, 6 K/min, and 12 K/min. Straight lines are the fit of these data using a first order kinetic Polanyi-Wigner law with a distribution of adsorption energy using a prefactor of 10$^{15}$ s$^{-1}$. (b) Search for the best value for the prefactor $\nu$: plot of the standard deviation of the three minimum occupied adsorption energies as extracted from the fits of the three TPD curves. The minimum of the curve give the best choice for $\nu$. (c) Adsorption energy distribution corresponding to 0.2 ML of CH$_3$CN on quartz using the best choice for the prefactor.  }
\end{figure*}


Since the energy is not expected to depend on the heating rate, the best choice for $\nu$ is the one for which the three distributions are found to be identical. Numerically, this can be achieved by searching the minimum standard deviation of the lowest populated adsorption energy for the three ramps as a function of $\nu$ (Fig. 5b). This gives an average value for the prefactor, which is the value we use for each species and each substrate in the following.

The results of this procedure are shown in  Figure 6 for different initial coverage of each isomer adsorbed on the three substrates. The value of the prefactor, the most probable adsorption energy (taken as the maximum of the distributions), and the size of the distributions defined as the width at half maximum  are
summarized for each adsorbate and each substrate in Table 2. From these values, a first conclusion is that, whatever the substrate, the adsorption energy of acetonitrile (CH$_3$CN) is found to be higher than that of methylisocyanide (CH$_3$NC). Desorption from quartz, on the one hand, and from water surfaces, on the other hand, show different behaviours, and are discussed separately in the following.
%
\begin{figure*}
\centering
\includegraphics[width=15cm]{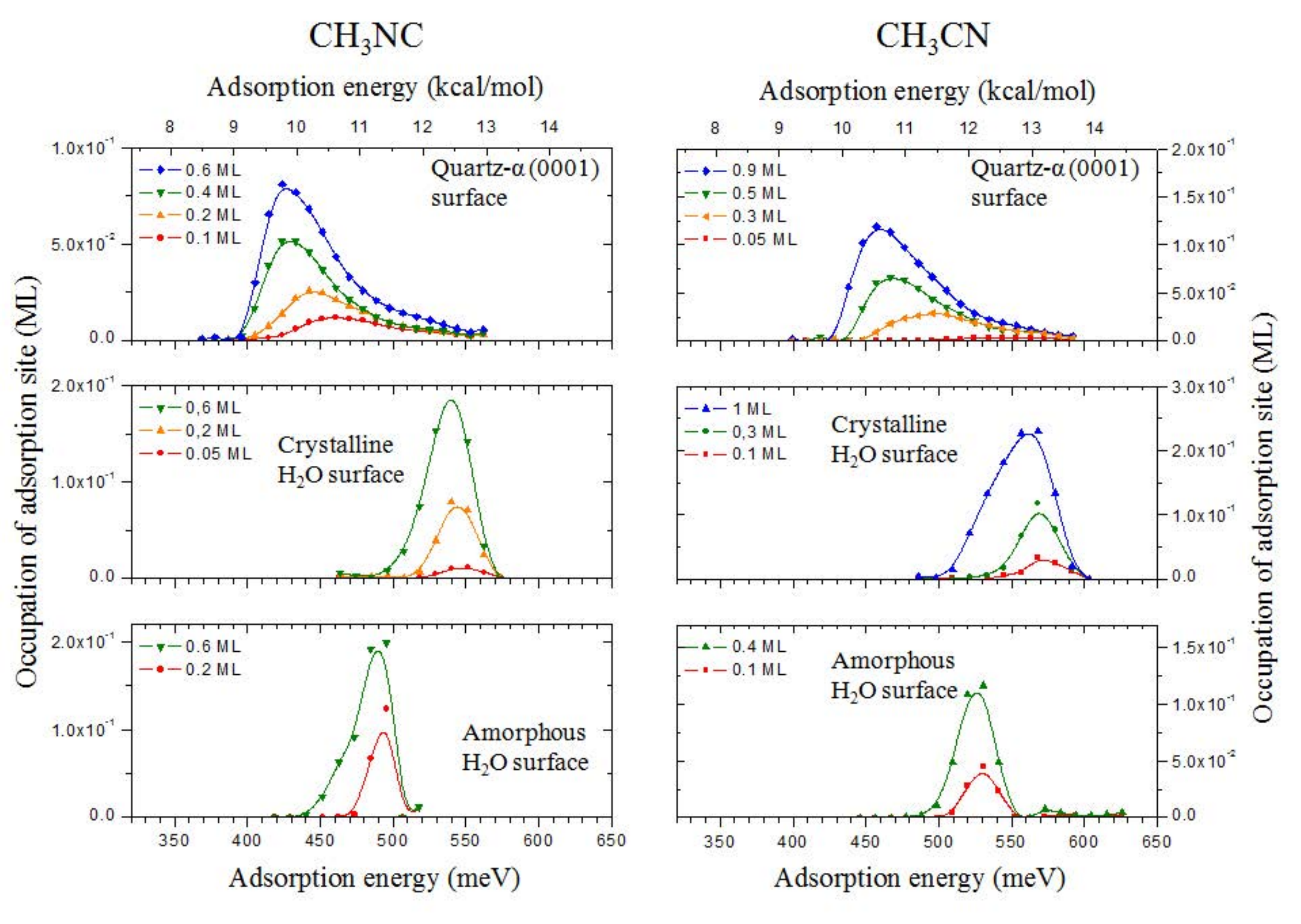} 
\caption{: Adsorption energy distributions for several initial submonolayer coverages of CH$_3$NC (left) and CH$_3$CN (right) adsorbed on $\alpha$-quartz (0001), crystalline water, and compact amorphous water surfaces. }
\end{figure*}

\begin{table*}
\begin{center}
\caption{Experimental prefactor $\nu$, most probable adsorption energy E$_{ads}$ and size of the adsorption energy distribution $\delta$E$_{ads}$ as taken as the full width at half maximum, for CH$_3$CN and CH$_3$NC adsorbed on $\alpha$-quartz (0001), and crystalline or compact amorphous water ice surfaces. Distinction is made between adsorption energy at high coverage (0.7-1 ML) and low coverage ($<$ 0.3 ML). \label{tab2} }
\begin{tabular}{ccccccc}     
\hline
\hline
\multirow{2}{*}{Substrates} & \multirow{2}{*}{Adsorbates} & \multirow{2}{*}{Prefactor $\nu$ (s$^{-1}$)} & \multicolumn{2}{c}{E$_{ads}$ (meV)} & \multicolumn{2}{c}{$\delta$E$_{ads}$ (meV)} \\
& & & 0.7 - 1 ML & $<$ 0.3 ML & 0.7 - 1 ML & $<$ 0.3 ML \\
\hline
\multirow{2}{*}{$\alpha$-quartz (0001)} &  CH$_3$NC & 3 $\times$ 10$^{17\pm0.5}$ & 430 & 460 & 50 & 70 \\
& CH$_3$CN & 1 $\times$ 10$^{17\pm0.5}$ & 460 & 495 & 60 & 75 \\
\multirow{2}{*}{Crystalline water ice} &  CH$_3$NC & 1 $\times$ 10$^{18\pm0.5}$ & 540* & 550* & 30 & 25 \\
& CH$_3$CN & 1 $\times$ 10$^{18\pm0.5}$ & 565* & 570* & 50 & 35 \\
\multirow{2}{*}{Amorphous water ice} &  CH$_3$NC & 5 $\times$ 10$^{16\pm0.5}$ & 490* & 490* & 25 & 20 \\
& CH$_3$CN & 2 $\times$ 10$^{17\pm0.5}$ & - & 530* & - & 30 \\
\hline
\end{tabular}

\end{center}
Values denoted with * should be considered with caution since the sublimation of the supporting water ice layer plays an important role in the observed desorption features. 
\end{table*}

In the case of adsorption on the quartz surface, two regimes can be distinguished. For initial coverage above 0.5 ML, the adsorption energy distribution is weakly dependent on the initial coverage. For initial coverage below 0.3 ML, the adsorption energy distribution significantly shifts towards higher energies with decreasing coverage. Indeed, at lower coverage, the molecules may diffuse during the warming up and fall in the tightest bounded adsorption sites. At higher coverage, as all these sites are populated, less bound sites are probed by the excess molecules. Experiments performed at very low coverage therefore enhance the role of the most bounded sites in the adsorption energy distribution, which may be defects on the surface. 

In the case of adsorption on water ices, conclusions are not straightforward and the results obtained should be considered with caution. The adsorption energy on crystalline ice is, for both isomers, found higher than in the case of adsorption on amorphous ice. However, it can be understood as follows: the adsorption energies of CH$_3$CN and CH$_3$NC  determined from the data treatment are very close to the adsorption energy of the underneath water ice itself, whose desorption has already started when the monolayer desorption of CH$_3$CN or CH$_3$NC is observed. The desorption energy of amorphous water is of 490 meV and that of crystalline water is found within 500-520 meV (Speedy et al 1996; Fraser et al 2010), which are very similar to the adsorption energy that we found for CH$_3$NC  on water. We believe the desorption we observe for CH$_3$CN and CH$_3$NC to be, at least partly, due to the desorption of the supporting water ice. Therefore the values that we obtain for the adsorption energies do not reflect the real interaction strength of the CH$_3$CN or CH$_3$NC molecules with the water ice surface, but suggest a lower limit instead. However, we still find a difference between CH$_3$CN and CH$_3$NC adsorption energies, which would not be the case if their desorptions were only due to the water sublimation. In particular, the adsorption energies for CH$_3$CN on water is found to be higher than that of water on water ice itself. Thus, the interaction of methylisocyanide and acetonitrile with water also plays a role in the measured adsorption energies, and we estimate that CH$_3$CN is bound with a higher energy to the amorphous or crystalline water surface than CH$_3$NC.


 \section{Theoretical approach to adsorption energies} 
 
 The adsorption energy is usually seen as a local property arising from the electronic interaction between a solid support and the molecules deposited on its surface. This interaction energy, E$_{ads}$ ,   is obtained  as 
 
  \[ E_{ads} = (E_{surf} + E_{mol}) - E, \]
 
 \noindent where here E$_{mol}$ is the energy of a single R-CN or R-NC molecule, E$_{surf}$ is the energy of the pristine surface of the substrate, and E is the  total energy of the [surface +  R-CN or R-NC] complex in which all entities are optimized in isolation. 
 
Two different ways of describing the solid surface are generally  considered, i.e. the cluster model and the  solid state periodic model. 

Representing a grain as a cluster seems to be a natural approach, but in this model, the surface is that of a molecular aggregate of limited dimension constrained by the number of molecules participating in the structure. This representation presents several drawbacks. For example,  the H$_2$O aggregates, when optimized, present very different surfaces (Maheshwary et al 2001; Buch et al 2004; James et al 2005) so that there is a complete loss of generality. 

For larger aggregates the calculations become rapidly intractable and over 200 molecules, the structure tends  to crystalline ice. 
Furthermore, the hydrogen bonding and van der Waals interactions are subject to basis set superposition errors (BSSE) that may add some bias to the calculated values (Boys \& Bernardi 1970).  
In the periodic representation there is no limitation to the surface size that is treated as the frontier of a solid of infinite dimensions. Furthermore, there is no BSSE when using plane waves for describing the electronic density.  

All calculations are performed by means of the Vienna Ab initio Simulation Package (VASP; Kresse \& Hafner 1993; 1994; Kresse \& {Furthm\"{u}ller 1996).
A plane wave basis set is used to represent the electron density. The core electrons are kept frozen and replaced by pseudo-potentials generated by the plane augmented wave method (PAW) as developed by Bl\"{o}chl (1994) and Kresse \& Joubert (1999). The grid is adapted to provide an equivalent treatment for all the unit cells whatever their size. Levering on preceding studies we employed the PBE generalized gradient approximation (GGA) functional (Perdew et al 1996) in the (PBE+D2) version corrected by Grimme et al. (2010)  for a better description of the long-range weak interactions. 
All calculations were carried out at this level of theory.
As periodic calculations were based on the replication of the unit cell along the three directions of space, the size of the unit cell is a critical parameter whose dimensions were determined here to avoid any spurious lateral and vertical interactions between successive slabs. The experience drawn from preceding studies (Lattelais et al. 2011; Lattelais et al. 2015) showed that a distance of $\sim$14 \AA between the adsorbed species is necessary to avoid the interactions generated by lateral translations. The energy minima were then obtained by full optimizations starting from most of the chemically reasonable positions (over atoms and depressions) and orientations of the adsorbates over the host surfaces.

   \subsection{Adsorption energies on silica}

 In this work, we used   $\alpha$-quartz (0001) in both TPD experiments and theoretical modelling; it is the most stable crystal allotrope of silica and the most stable surface (Leven et al 1980). The active surface was hydroxylated,  which means that there are no free oxygen atoms left on the surface.  It is consistent with the calculations by Goumans et al. (2007), which showed that, once hydroxylated, the  $\alpha$-quartz (0001) surface was more stable than the raw surface obtained by the simple cut of a single crystal.                                
 
 The $\alpha$-quartz (0001) surface model  was constructed from an appropriate cut of
the top three silicon layers of the experimental crystal structure, using the experimental lattice parameters. The unit cell is then re-optimized in the hydroxylated form .  
 In the exploitation phase, we take a $4\times4$ basal surface, reaching a $19.66\times21.62$ \AA$^2$ area with a vacuum 15\AA high to be safe from any unwanted lateral or vertical interactions in probing the adsorption sites. 
The two bottom layers of Si and surrounding oxygen atoms
were constrained to maintain the slab geometry.

The CH$_3$CN and CH$_3$NC isomers present close adsorption energies on silica. Both molecules are rigid enough to stay linear and both are tilted in entirety over the hydroxylated surface with two of the CH bonds pointing towards the oxygen lone pairs of the surface. The N and C extremities of the CN, respectively NC, bonds interact  (N more  than C) with the dandling OH bonds of the surface, strongly enough to induce changes in the geometries of the hydroxyl surface arrangement (Table 3(top)). By contrast, the HCN and HNC isomers have well-separated adsorption energies on silica. They remain quasi-linear,  where the CH or NH bonds are oriented  with the hydrogen atom towards the electrons lone pair of a surface oxygen (Table 3(bottom)). That HCN is less attached to the solid surface than HNC  is a clear illustration of the different polarities of the  H-C and H-N bonds, leading  the former to be engaged in a weaker hydrogen bond  with the oxygen lone pairs of the surface. 

Comparing the two couples of isomers, we emphasize two points. First, it can be seen that the difference in the adsorption energies of HCN and HNC reaches $\sim$200 meV whereas it is only of $\sim$40 meV for the CH$_3$-substituted species.   Second,  the order of increasing adsorption energies is reversed between the HCN and HNC and CH$_3$CN and CH$_3$NC couples of isomers.

\begin{table}
\begin{minipage}{\columnwidth}
\caption{Views of the most stable geometries of the adsorbed CH$_3$CN and CH$_3$NC, and HCN and HNC couples of isomers  on $\alpha$-quartz (0001).  Adsorption energies E$_{ads}$  are given in meV and kcal/mol.}
\centering 
\begin{tabular}{@{}cc@{}}
\hline\hline 
 
CH$_3$CN           & CH$_3$NC                        \\ 
 
\hline
 \\
\includegraphics[scale=0.33]{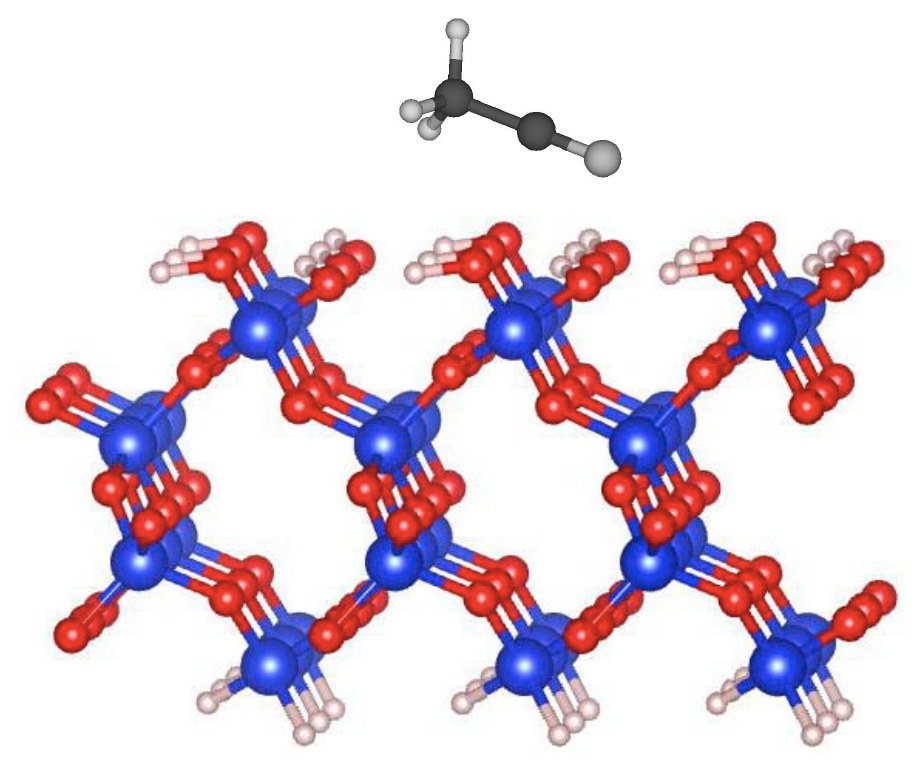} & 
\includegraphics[scale=0.33]{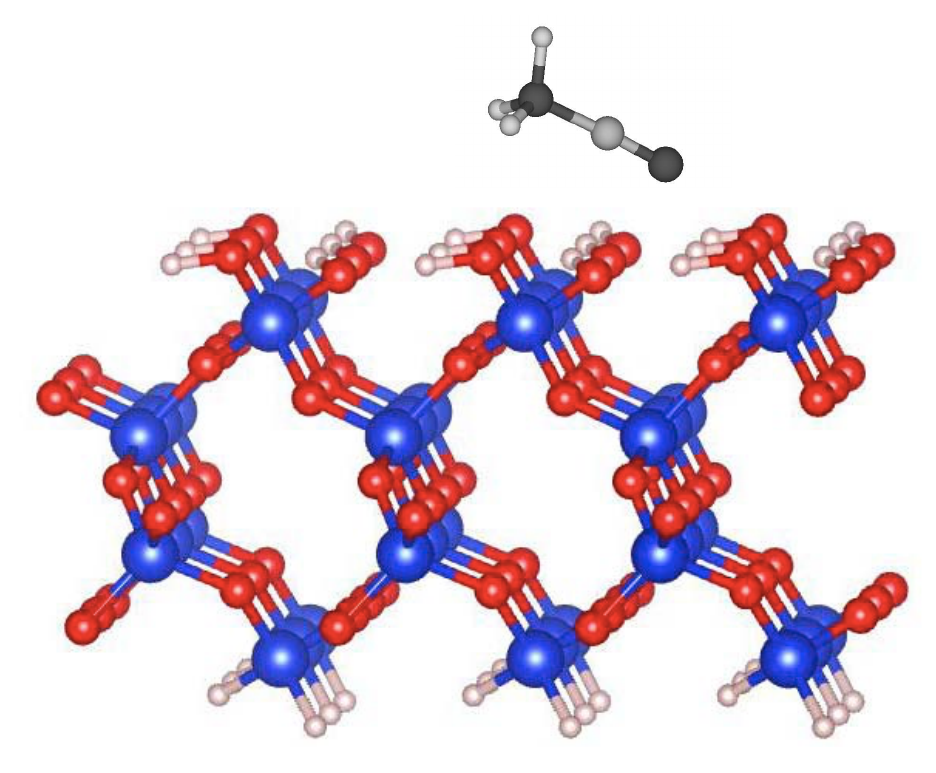}  \\
\\
E$_{ads}$ = 460 meV & E$_{ads}$ = 414 meV \\
         10.61 kcal/mol     &             9.55 kcal/mol   \\                 
 
\hline

 HCN                  &  HNC  \\
 
 \hline
\\ 
\includegraphics[scale=0.33]{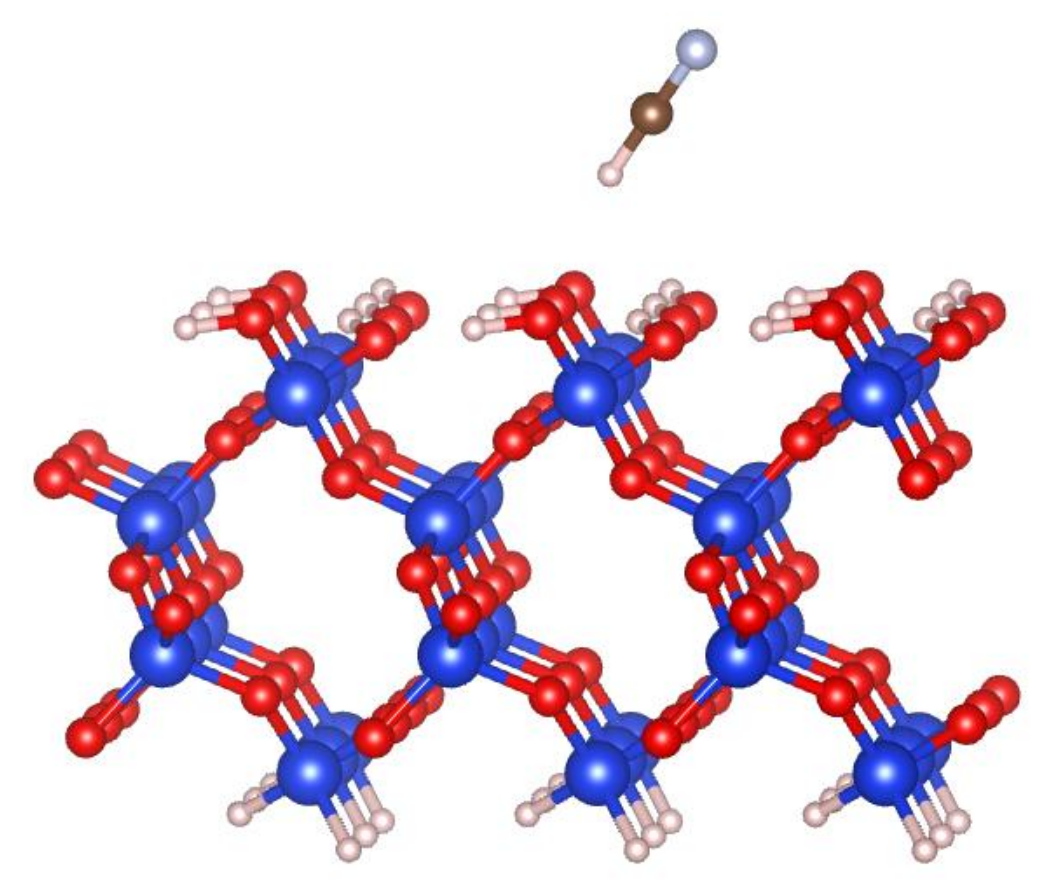} & 
\includegraphics[scale=0.33]{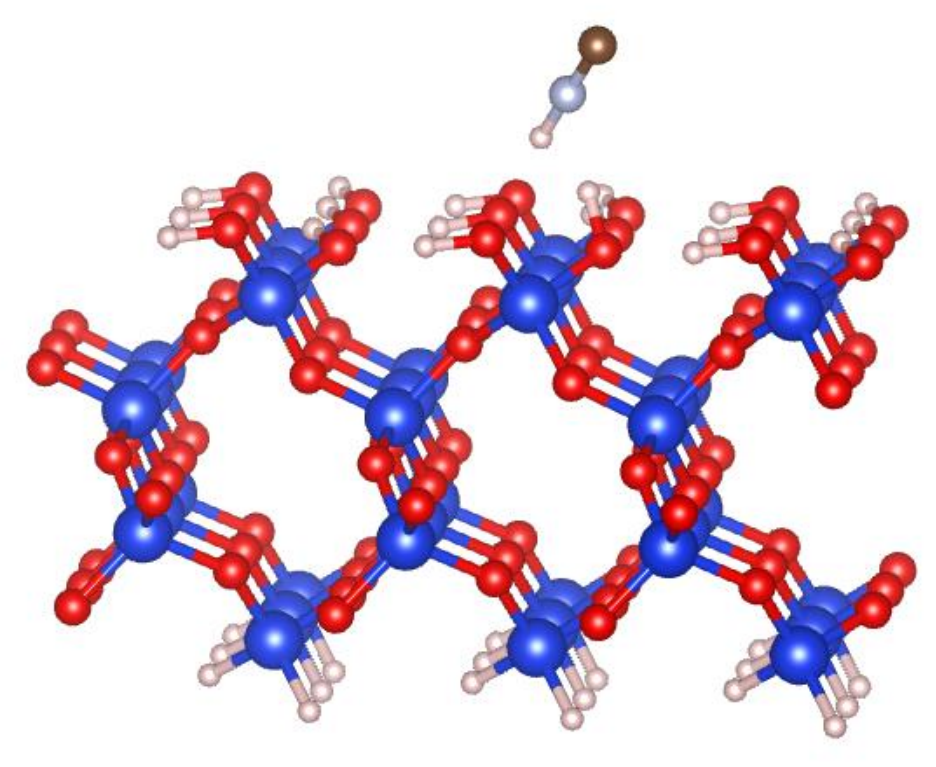}  \\
\\
E$_{ads}$ =  212 meV & E$_{ads}$ =   421 meV \\
         4.89 kcal/mol     &             9.72 kcal/mol   \\                 

\hline
\end{tabular} 
\end{minipage}
\end{table}

  \subsection{Adsorption energies on crystalline water ice}

The structure of ice has been the centre of a number of computational studies in the past (Casassa \& Pisani 2002; Kuo \& Singer 2003; Hirsch \& Ojamae 2004; Casassa et al. 2005). Among the different structures possible, we focused on the apolar variety of crystalline ice because only apolar structures can generate slabs that are stable, reproduce the bulk properties, and have a balanced distribution of alternate hydrogen and oxygen sites at their surfaces (for a complete discussion, see Casassa et al. 2005). 
 
 The unit cell used to construct the slab is that of hexagonal ice {\it Ih} containing two bilayers. Following the same protocol as exposed above, this unit cell is re-optimized. In the exploitation phase, we take a basal surface reaching a  $14.29\times19.64$ \AA$^2$ area with a vacuum 15\AA high to be safe from any unwanted lateral or vertical interactions in probing the adsorption sites. 

Table 4(top) shows that the CH$_3$CN and CH$_3$NC isomers adsorb on the surface by two types of hydrogen bonds: one between an oxygen atom from the ice surface and two H atoms of the methyl group of  the adsorbed isomer;  the other  between an H atom of a OH bond from the surface and the N or C atom at the other extremity of the adsorbate molecule. Both adsorption energies are practically identical, but once more the same trend is observed 
with E$_{ads}$(CH$_3$CN) $>$ E$_{ads}$(CH$_3$NC).

In the case of HCN and HNC, it is the same situation as for the hydroxylated silica surface; namely, HNC is much more strongly attached to the surface than HCN with a 180 meV energy difference. The structures of these triatomic species are more flexible and they are no longer strictly linear (Table 4, bottom). If HCN, known to be rigid in isolation, remains close to linear, HNC is more flexible and significantly bent 
with  $\langle$HNC$\sim$160 $\deg,$ which allows  a much better adaptation to the  ice giving  rise to a stronger hydrogen bond with the surface.

Comparing the behaviour of both isomers couples leads exactly to the same conclusion as in the case of the silica surface.

   
\begin{table}
\begin{minipage}{\columnwidth}
\caption{Views of the most stable geometries of the adsorbed CH$_3$CN and CH$_3$NC, and HCN and HNC couples of isomers  on apolar water ice.  Adsorption energies E$_{ads}$  are given in meV  and kcal/mol.}
\centering 
\begin{tabular}{@{}cc@{}}
\hline\hline 
 
CH$_3$CN           & CH$_3$NC                        \\ 
 
\hline
 \\
\includegraphics[scale=0.17]{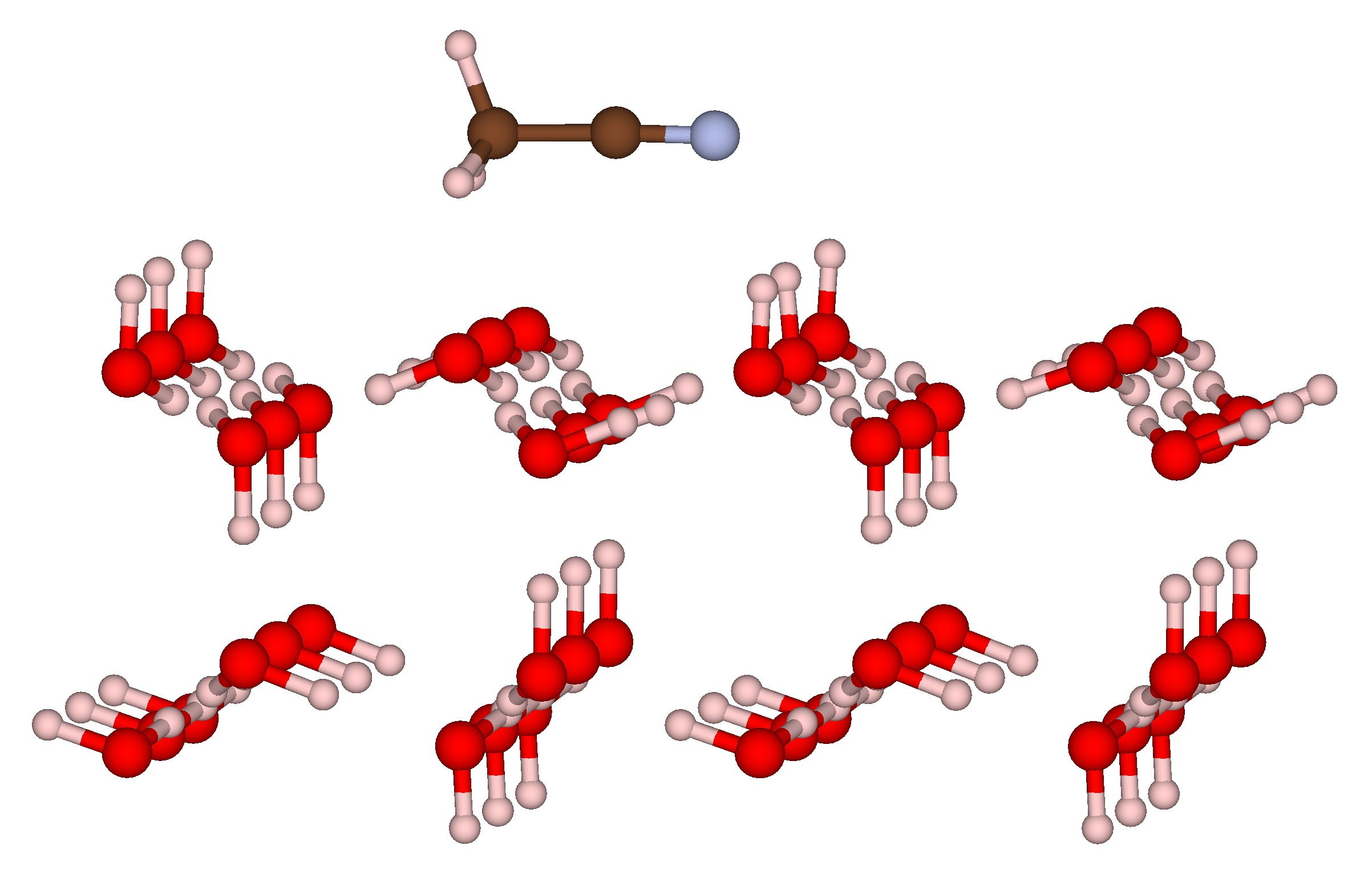} & 
\includegraphics[scale=0.13]{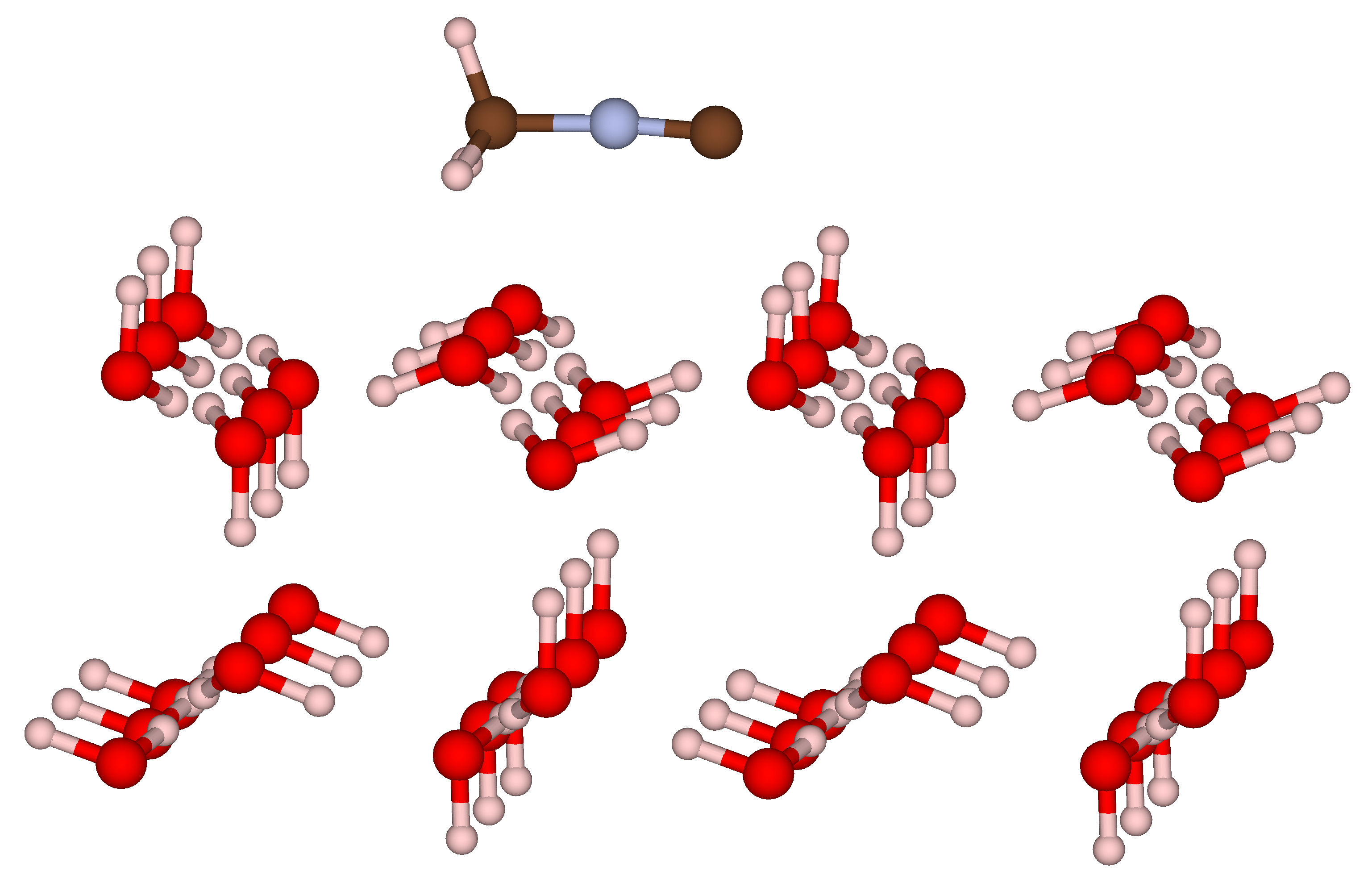}  \\
\\
E$_{ads}$ =  558 meV  & E$_{ads}$ = 545 meV  \\
         12.87 kcal/mol     &             12.57 kcal/mol   \\                 
 
\hline 

HCN     &  HNC   \\ 

\hline
 \\
\includegraphics[scale=0.14]{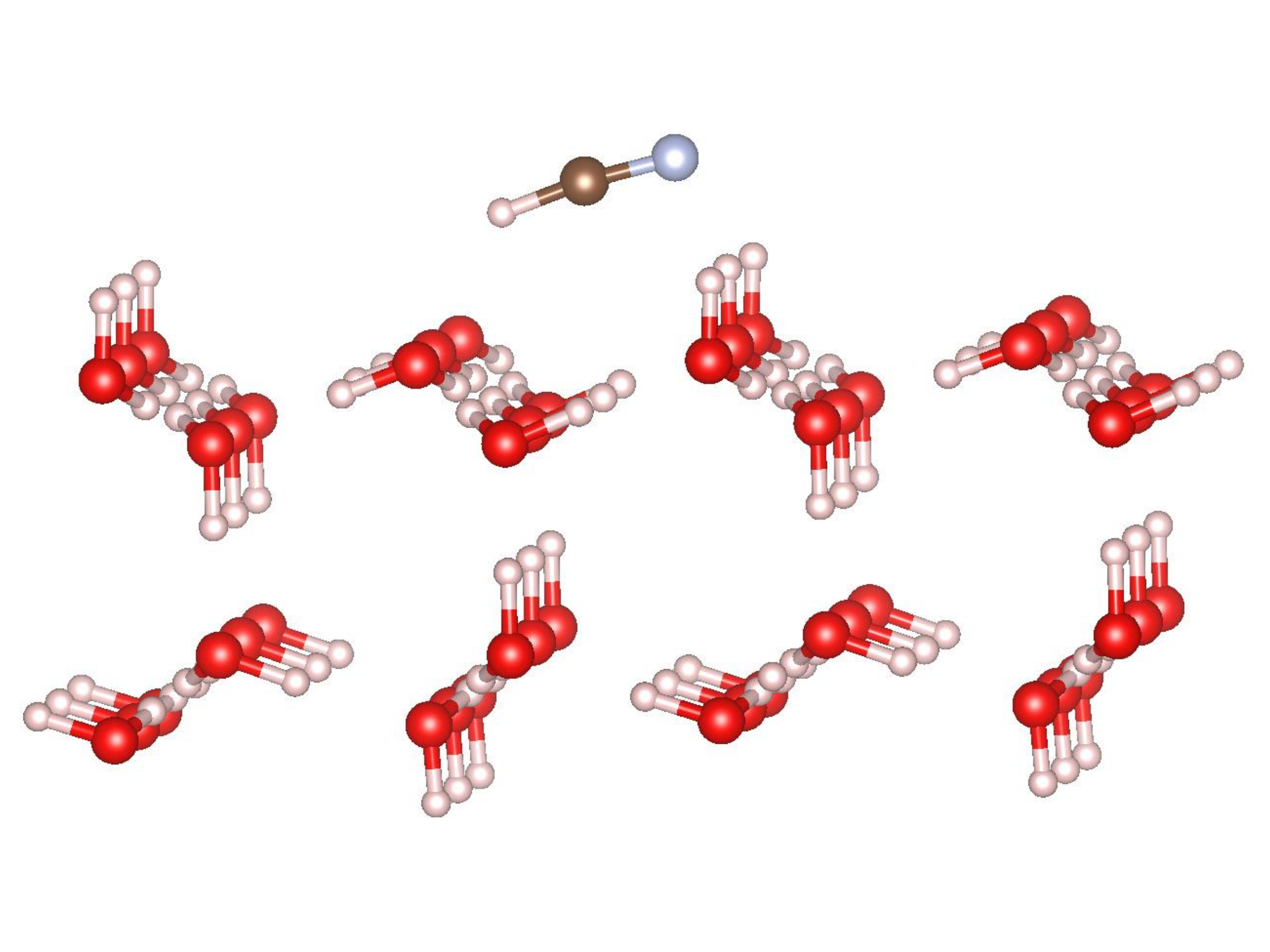} & 
\includegraphics[scale=0.14]{Fig-Glace-HCN.pdf}
\\
E$_{ads}$ = 601 meV  & E$_{ads}$ = 776 meV  \\
         13.87 kcal/mol     &             17.91 kcal/mol   \\                 

\hline
\end{tabular} 
\end{minipage}
\end{table}

   \section{Discussion and final remarks}

The adsorption energies of the two isomers CH$_3$NC and CH$_3$CN on two different astrophysically relevant surfaces were studied, both experimentally, using the TPD technique in a state-of-the-art ultra-high vacuum system, and theoretically, by means of first principle  periodic DFT calculations.  This concerted approach allows for determining values of adsorption energies, obtained with two independent methods, whose reliability is supported by thorough cross-checking  between experiments and numerical simulations. 
The validity of such an approach has already been illustrated in the study of the adsorption of complex organic molecules on water ice (Lattelais et al 2011).
In the present work, the adsorption of the two nitrile and isonitrile isomers has been studied on a silica surface, as a model for bare interstellar  grain surface, and on a water ice surface, as water is the most abundant constituent of the icy mantles recovering the dust grains. 


Three main conclusions can be drawn from the results. 

\begin{enumerate}
\item Comparing the computed and experimental values summarized in Table 5, shows clearly that experiments and theoretical calculations for adsorption energies give very similar results in all cases. The case of the silicate surface case, case which has not been tested in previous studies, is particularly demonstrative.   We obtain a remarkable  agreement between experimental values (430 and 460 meV) and theoretical values (414 and 460 meV) for CH$_3$NC and CH$_3$CN adsorption, respectively. 
 In the case of the crystalline water ice surface, we also found the values to be very close between  experiment and theory.  This further confirms the validity of the method we use for adsorption energies determination.\\
That the underlying water ice desorbs simultaneously with CH$_3$NC or CH$_3$CN makes it plausible that the experimental values obtained could  represent a lower limit  for the adsorption energy of CH$_3$NC and CH$_3$CN on crystalline water. In this case only numerical simulations allow  for unbiased determination of  the adsorption energies - 545 and 558 meV for  CH$_3$NC or CH$_3$CN, respectively - are found very close to the water adsorption energy on crystalline water ice. It is what was expected from the experimental results. Finally, the experiments realized on amorphous water ice show that a small part of the adsorbed molecules desorb through a volcano effect, meaning that some diffusion took place during the warming up and that some of the CH$_3$CN or CH$_3$NC molecules get trapped in the corrugation of the amorphous water bulk. This trapping ability was not studied in detail here, but may also play a role in the surface-to-gas abundances of these species in presence of water-rich icy coating in the ISM. Besides this inclusion effect,  interestingly, we found the adsorption energies on the amorphous water surface  to be very close to those on the crystalline ice surface, both experimentally and theoretically.
\smallskip 
\smallskip 

\item Differences in adsorption energies between the nitrile CH$_3$CN and  the isonitrile CH$_3$NC are almost the same for the two surfaces considered.
However, the adsorption energies of CH$_3$CN are always found to be a little higher than those of CH$_3$NC by 20 to 40 meV, both experimentally and theoretically. This cannot be explained by different values of the dipole moments since both are close to $\sim$4  Debye (4.0 and 3.9 for CH$_3$CN and CH$_3$NC, respectively), but this can instead be due to a slightly stronger ability of CH$_3$CN to perform H-bonds  because of a local excess of positive charge on the hydrogens of the methyl group. \\
Such a similar behaviour of the two isomers might also be a reason why their abundance ratio  is  relatively constant in many different ISM regions. It is true  from the colder regions, where the interaction of the species is more likely to occur with the icy mantle, mainly composed of water, to warmer regions where the adsorption and re-adsorption on the naked  silicated grain surface may begin to play an important role.  Since the difference in adsorption energies does not depend on the substrates, the differential desorption effect between isomers on the gas-phase CH$_3$CN to CH$_3$NC abundance ratio should be the same whatever  the  region considered in the ISM. 

\smallskip 
\smallskip 

\item The situation is different with HCN and HNC whose adsorption energies were computed for the same surfaces (no experimental study has been attempted in this case). The values of 212 and 421 meV for HCN and 601 and 776 meV for HNC on hydroxylated silica and crystalline ice, respectively, (see lower panels of Tables 3 and 4)  show that the intrinsically less stable isomer HNC is more strongly adsorbed than HCN on all surfaces, contrary to the general trend observed for the methyl-substituted compounds. \\
In addition, the adsorption energies of HNC are larger than those of HCN by a similar amount (about $\sim$200 meV for silica and water ice, i.e. for two substrates favouring the formation of OH bonds. A similar result, with the same origin,  was obtained in a preceding study on the HCNO and HNCO couple of isomers whose attachments to the silica surface were found to be of $\sim$130 and 240 meV, whereas they are of $\sim$460 and 600 meV on the ice,  respectively  (Lattelais et al. 2015). \\
In any case, the different adsorption behaviour between HCN and HNC  on naked grain surfaces  or on water ice would likely induce different abundance ratios in different environments, which is well in line with  their observed fluctuating abundance ratios in different regions of the ISM.

\end{enumerate}

\begin{table*}
\begin{center} 
\centering
\caption{Computed and experimental adsorption energies (meV) of CH$_3$CN and CH$_3$NC on $\alpha$-quartz (0001) and H$_2$O surfaces.}
\label{table:1b}
         
\begin{tabular}{l c c c c  }     
\hline\hline       
                                & \multicolumn{2}{c}{CH$_3$CN}   & \multicolumn{2}{c}{CH$_3$NC}   \\ 
Surface                         & Computed & Experimental & Computed & Experimental       \\
\hline                                        
 & & \\

 $\alpha$-quartz  (0001)        &       460     & 460 $\pm$ 30 & 414 & 430 $\pm$ 25        \\
 apolar crystalline ice {\it Ih}                        &       558     &  565 $\pm$ 25 * &        545 & 540 $\pm$ 15 *   \\
 compact amorphous ice      & - & 530 $\pm$ 15 * & - & 490 $\pm$ 10 * \\
  
   \hline                  
\end{tabular}
\end{center}
Experimental values denoted with * should be considered with caution since the sublimation of the supporting water ice layer plays an important role in the observed desorption features. 
\end{table*}


\begin{acknowledgements}
  This work was supported by CNRS national programme PCMI (Physics and
  Chemistry of the Interstellar Medium) and COST Action CM 1401, "Our astrochemical history". We thank the UPMC labex MiChem for providing Ph.D. financial support for M. Doronin and for financial help on the design of the experimental setup. 
  \end{acknowledgements}

\end{document}